# Imaging Simulation Of A Dual-Panel PET Geometry With Ultrafast TOF Detectors

Taiyo Ishikawa, Go Akamatsu, Hideaki Tashima, Fumihiko Nishikido, Fumio Hashimoto, Ryosuke Ota, Hideaki Haneishi, Sun Il Kwon, Simon R. Cherry, and Taiga Yamaya

***Abstract*—In positron emission tomography (PET), time-of-flight (TOF) information localizes source positions along lines of response. Cherenkov-radiator-integrated microchannel-plate photomultiplier tubes have achieved 30 ps TOF resolution, demonstrating cross-sectional imaging without reconstruction. Such ultrafast TOF detectors would free PET from conventional ring geometries. Therefore, this study aimed at investigating imaging characteristics of a dual-panel PET with ultrafast TOF detectors using Geant4 simulation. Two detector panels (137 × 137 mm$^2$), which consisted of 5.0 mm-thick bismuth germanate pixelized crystals with a 5.75 mm pitch, were placed face-to-face at a 300 mm distance. Imaging characteristics with various TOF resolutions from 30 to 90 ps were evaluated. Because degraded efficiency may cancel TOF gain in image quality, detection efficiency was also parameterized by reducing coincidence counts. Data acquisitions for a numerical multi-rod and uniform phantom (21 MBq) and a modified NEMA NU2 image quality phantom were simulated for 600 s. Results of the maximum likelihood expectation maximization (MLEM) reconstruction were compared with those of a backprojection (i.e., no reconstruction). The dual-panel PET required a 40 ps TOF resolution to have a similar spatial resolution to that of a non-TOF ring PET (300 mm in diameter) for the same detection efficiency. TOF showed benefit in the reconstruction of image quality phantom with 40% efficiency, and the image noise with 20% efficiency at 30 ps TOF was similar to the complete efficiency at 40 ps TOF. MLEM provided better imaging performance than backprojection, even at 30 ps TOF. The feasibility of the proposed dual-panel PET was shown.**

## I. INTRODUCTION

IN positron emission tomography (PET), time-of-flight (TOF) in the coincidence detection localizes the source position along a line of response (LOR) [1], [2]. Researchers have continued to improve TOF resolution [3] and it has reached around 200 ps, corresponding to 3 cm spatial resolution, in a state-of-the-art commercial clinical PET systems [4]. On a laboratory bench scale, a pair of Cherenkov-radiator-integrated microchannel plate photomultiplier tubes (CRI-MCP-PMT) has achieved a TOF resolution of ~30 ps [5]; the TOF resolution itself has become almost equal to the spatial resolution of clinical PET systems [6]. Kwon *et al.* [7] have demonstrated cross-sectional imaging by estimating the source point directly from TOF information obtained with a pair of detectors, without rotation and reconstruction. This ultrafast TOF resolution is expected to free PET from a conventional ring-shaped detector geometry, which is expected to open up new applications of PET.

Among such non-ring geometries, a dual-panel or partial ring geometry has been investigated for almost two decades [8], [9], [10], [11], [12], [13], [14]. Some organ-dedicated PET systems have been developed with the expectation of improved spatial resolution by placing detectors closer to each other [10], [15], [16]. In addition, a new concept named walk-through PET, which utilizes an imager with a standing position to realize high throughput and low patient burden, has also been proposed [17]. The dual-panel geometry ensures a large open space around the patient, which is appropriate for PET-guided surgery [18], [19] and in-beam particle therapy monitoring [20]. For such applications, compact detector panels should be positioned at a sufficient distance from each other to allow the patient's body to be placed between them. In these systems, however, a limited projection view causes artifacts in the reconstruction [12]. TOF is expected to reduce these artifacts, but the currently available TOF resolution of around 200 ps is insufficient for the dual-panel PET systems for the PET-guided surgery and beam monitoring applications.

It was demonstrated that 30 ps TOF resolution enabled cross-sectional imaging in the bench-top prototype [7]. However, the required TOF resolution and detection efficiency for the extension to a clinical system have yet to be properly investigated. The detection efficiency of the bench-top prototype [7] was low because of the low density (3.9 g/cm$^3$) and low refractive index (n ≌ 1.5 for 400 nm) of the lead glass used as the Cherenkov radiator [21]. On the other hand, bismuth germanate (BGO) crystal is attracting attention as a Cherenkov and scintillation emitter [22], [23], [24], [25], [26], [27], [28] because its higher density (7.13 g/cm$^3$) [29] and higher refractive index (n ≌ 2.2 for 400 nm) [30] than those of the lead glass is expected to improve the detection efficiency. The scintillation emission of BGO provides energy

This work did not involve human subjects or animals in its research.

Taiyo Ishikawa, Fumio Hashimoto, Hideaki Haneishi, and Taiga Yamaya are with Chiba University, Chiba 263-8522, Japan (e-mail: t_ishikawa@chiba-u.jp). Taiyo Ishikawa, Go Akamatsu, Hideaki Tashima, Fumihiko Nishikido, Fumio Hashimoto and Taiga Yamaya are with the National Institutes for Quantum Science and Technology (QST), Chiba 263-8555, Japan (e-mail: yamaya.taiga@qst.go.jp). Fumio Hashimoto and Ryosuke Ota are with Hamamatsu Photonics K.K., Hamamatsu 434-8601, Japan. Sun Il Kwon and Simon R. Cherry are with the University of California, Davis, Davis, CA 95616, USA information, which is required to reject scatter coincidence events. Nevertheless, detecting and triggering on Cherenkov photons is challenging because only about 17 Cherenkov photons are produced [23], which is still a smaller number than scintillation photons (8200 photons/MeV) [31]. The fraction of coincidence events that both single events triggered by Cherenkov photons was several tens of percent depending on the measurement condition [24], [28]. The low detection efficiency may cancel the benefit of better TOF in terms of image noise. In addition, although the previous research claimed that reconstruction is unnecessary, the image quality of direct imaging has yet to be compared with that of iterative reconstruction.

This study focused on investigating the imaging characteristics of a dual-panel PET with ultrafast TOF resolution, in anticipation of the expansion of the setup of [7] to a human-sized clinical system. We quantitatively evaluated the spatial resolution, image noise, and contrast to determine the required TOF resolution.

## II. Materials and Methods

### A. Geometry

Fig. 1 shows the detector geometries of the proposed dual-panel and a ring PET. The ring PET was modeled as a reference for spatial resolution. The dual-panel PET had two detector panels (137 mm × 137 mm) placed face-to-face at a 300 mm distance. The ring PET had five detector rings with a 300 mm diameter and 137 mm axial length.

The dual-panel PET was composed of 50 detectors, and the ring PET was composed of 160 detectors. The detectors were composed of 4 × 4 arrays of pixelized bismuth germanate (BGO) crystals, each sized as 5.75 × 5.75 × 5.0 $mm^3$. This choice was based on multi-anode CRI-MCP-PMTs (5.75 mm readout pitch) [32] with a BGO window as a future possible extension of currently commercially available multi-anode MCP-PMTs (R10754-07-M16, Hamamatsu Photonics K. K.) [33].

### B. Simulation

We modeled the dual-panel and ring geometries in the Geant4 Monte Carlo simulation [34]. We did not implement optical simulation this time. We assumed that Cherenkov photons provided detection timing information, and scintillation photons provided position and energy information in the dual-panel PET, based on the properties of BGO [22].

The decay process of a positron emitter was approximated by directly generating a positron without kinetic energy. Generated positrons were annihilated with electrons in the air. The effects of positron range and photon non-colinearity were neglected.

During the simulation, position, energy, and detection timing information were obtained for each energy deposition. In the case of multiple interactions in the detector, the crystal with the maximum energy deposition was treated as the detection position. Deposited energy was summed for each array, and energy resolution was applied using a Gaussian function with 20% full width at half maximum (FWHM) [35]. We used a 400 – 600 keV energy window.

The detection timing was calculated from the interaction time at the detection position and the distance between the interaction position with the maximum energy deposition and the crystal bottom surface. The photon travel time was added to the interaction time, based on the assumption that the detected first photon traveled linearly to the photocathode without reflection or scattering. For simplicity, the speed of light in the BGO crystals was calculated assuming a photon with wavelength 490 nm and a refractive index of 2.15 [30]. Coincidence detection information was recorded as list-mode data.

The TOF resolution was modeled with a Gaussian function. A random number generated from the Gaussian function was added to TOF information, which was the difference between the two detection timings of the list-mode data. The detection timing uncertainty before TOF resolution modeling should be considered because the detection timing fluctuates depending on the depth of interaction (DOI). We acquired the timing uncertainty from data acquisition of a point source positioned at the middle point of the detector panels. As a result, the timing uncertainty was 20.1 ps FWHM. We set the width of the Gaussian function to achieve targeted TOF resolutions considering this timing uncertainty. In the Cherenkov/scintillation hybrid measurement with BGO crystals, the TOF resolution is well modeled by multiple Gaussian distributions with different FWHMs [25]. In this study, we assumed the dual-panel PET could reject single events where no Cherenkov photons were detected, and coincidence events were triggered by Cherenkov photons for TOF resolution modeling. On the other hand, the ring PET was a non-TOF system, and coincidence events were triggered by scintillation photons.

Cherenkov photons produced via the interaction of annihilation radiation with the BGO crystals are highly variable and may not be detected [23]. We simulated the reduction in detection efficiency by randomly thinning out the list-mode data. The ratio of events remaining after thinning was defined as the relative detection efficiency (RDE). In other words, RDE was the hypothetical ratio of the Cherenkov-Cherenkov events among all the coincidences.

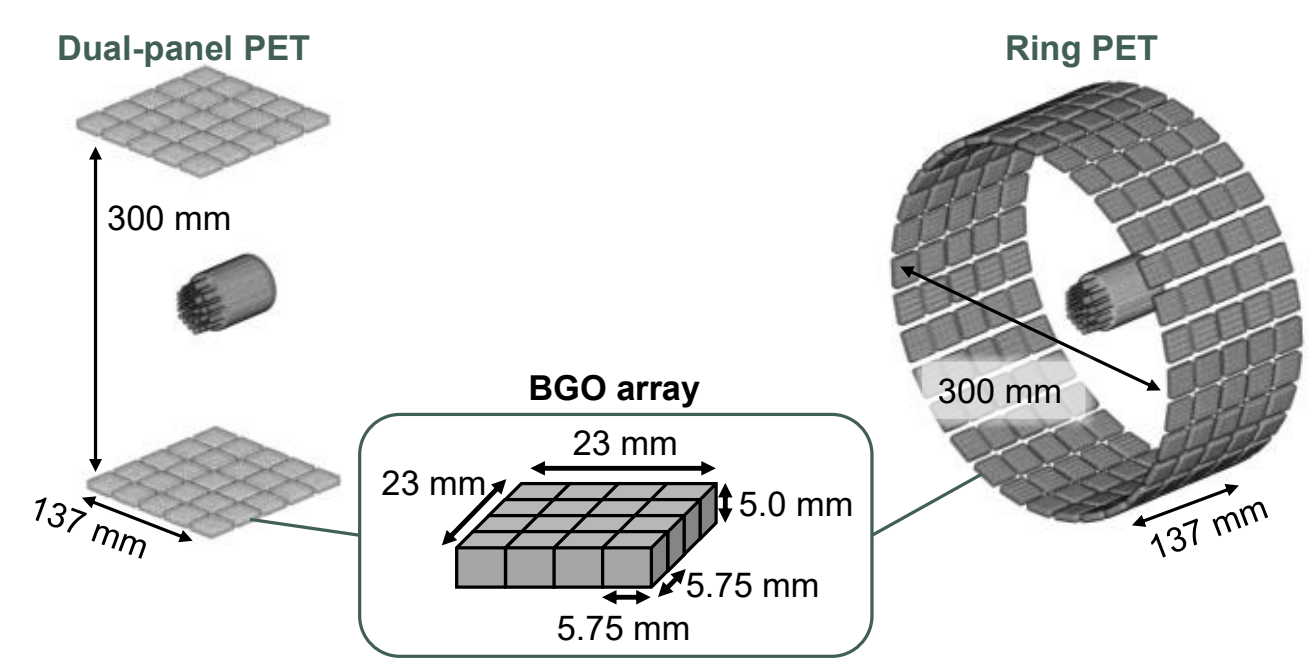

**Fig. 1.** Simulated geometries of the dual-panel PET and the ring PET. The detector had 4 × 4 array of BGO crystals, each sized 5.75 × 5.75 × 5.0 $mm^3$. The dual-panel PET had 50 detectors (5 × 5 detectors per panel), whereas the ring PET had 160 detectors (32 detectors per ring).

### *C. Numerical Phantoms*

We defined a numerical multi-rod and uniform phantom composed of a multi-rod area and a uniform area to evaluate the spatial resolution and image noise in a single measurement (Fig. 2). The multi-rod area consisted of clusters of rods, 15 mm in height and 6, 5, 4, 3, 2.2, or 1.6 mm in diameter. The uniform area consisted of a cylinder, 50 mm in height and 45 mm in diameter. Only the radioactivity distribution was defined, and no attenuating material was simulated. We simulated 600-s of data acquisition for the phantom with 21 MBq activity.

We also defined a numerical image quality phantom (Fig. 3) by modifying the NEMA NU2 image quality phantom [36] to match the dual-panel geometry. The phantom was composed of a water cylinder with a 136 mm diameter and an axial length of 48 mm. The cylindrical background activity area had a 130 mm diameter and an axial length of 42 mm. Hot spheres with 22, 17, 13, and 10 mm diameter were placed in the central slice. A cold cylinder with a 30 mm diameter and an axial length of 42 mm was centered. The activity concentration of the background was 5.3 kBq/ml, and the hot spheres had a four times higher activity concentration than the background. The half-life was set assuming the use of $^{18}$F. The data acquisition time was 600 s.

### *D. Image Reconstruction*

We used list-mode maximum likelihood expectation maximization (MLEM) and backprojection (i.e., no reconstruction) for image reconstruction. We used in-house code [37], and we implemented detector response function modeling with a TOF response function for projector and backprojectors in the MLEM. The detector response function in the direction perpendicular to the LOR was modeled as a Gaussian function [38] for the distance between each LOR and the center of each voxel with 2.5 mm FWHM. The TOF response function was modeled by convolving a rectangular function of TOF bin size with a Gaussian function having the FWHM of TOF resolution [39]. Let the $l$-axis be a line along a LOR. The center where a coincidence event is recorded is given by

$$l_0 = \frac{ct}{2}, \quad (1)$$

where $c$ is the speed of light, and $t$ is the difference of arrival time. The probability density function along the LOR was given by

$$p(l) = \frac{1}{D} rect\left(\frac{1}{D}(l - l_0)\right) * \frac{1}{\sqrt{2\pi}\sigma} \exp\left\{-\frac{l^2}{2\sigma^2}\right\}, \quad (2)$$

where $D$ is the TOF bin size, which was 1.2 mm, and $\sigma = \frac{\text{FWHM}}{2\sqrt{2\ln 2}}$ is the standard deviation. We should note that only the MLEM without TOF information was applied for the ring PET because we used it as a reference for the spatial resolution. The voxel size of the reconstructed images was 1.0 × 1.0 × 1.0 mm$^3$.

A direct method was used for normalization. A hollow ring phantom was used for the ring PET. The hollow ring consisted of a positron distribution of 1 MBq, 2 mm in thickness, 288 mm in diameter, and 142 mm in axial length. The data acquisition time was 64800 s. For the dual-panel PET, a flat plate-shaped positron distribution of 1 MBq with a surface of 142 × 142 mm$^2$ and a thickness of 2 mm was set. The phantom was placed in the middle between the two detector panels and parallel to them. The data acquisition time was 21600 s.

We used the same backprojector as MLEM for backprojection. TOF response function modeling in backprojection will blur reconstructed images. A most-likely-point backprojector was also used to suppress the blurring. We assumed that the most likely point was $l_0$ in (1). To use the same backprojector as the TOF response function, the rectangular function was convolved with a Gaussian function with 1.0 ps FWHM (0.15 mm spatial information). Dividing the backprojection image with a GSI, which was the backprojection of all the normalization factors, enabled the sensitivity correction.

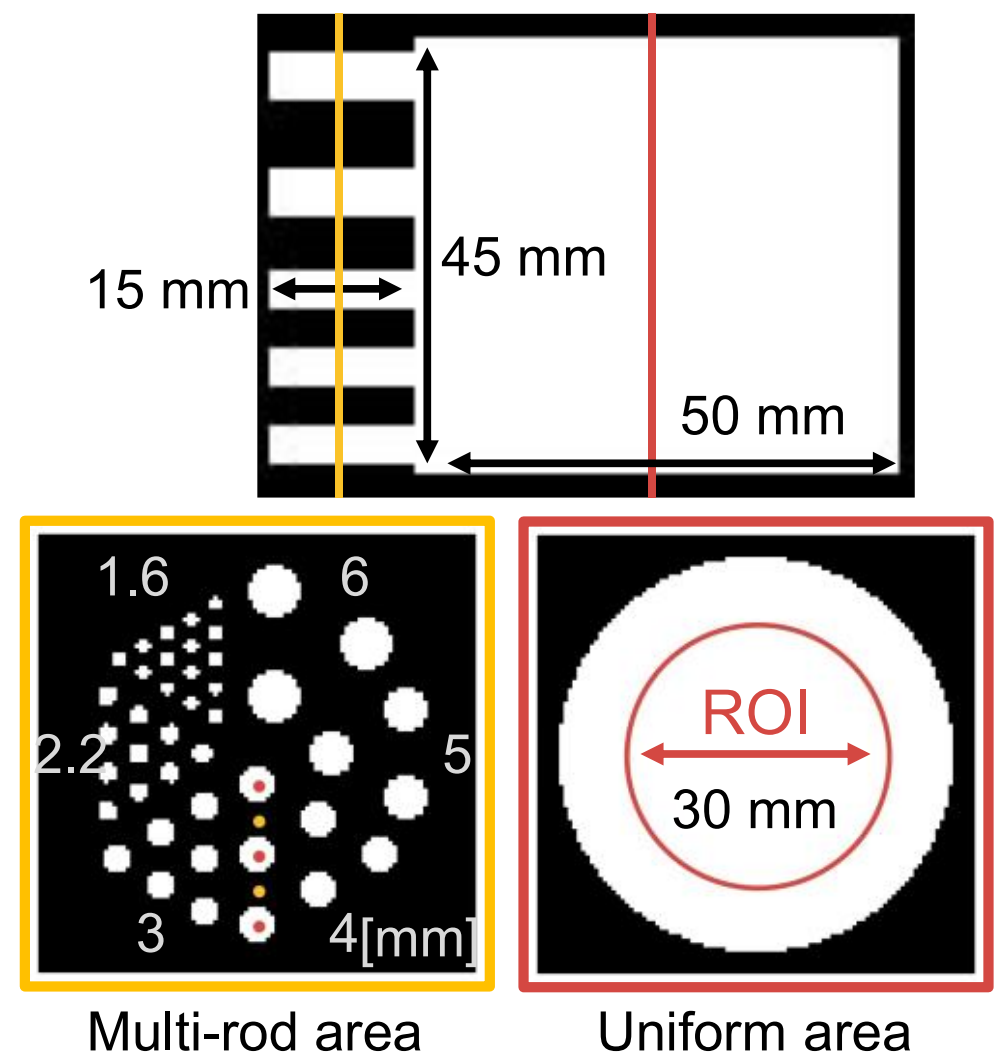

**Fig. 2.** Activity distribution of the numerical phantom. The phantom did not interact with annihilation radiation. The spatial resolution was evaluated through the percentage peak-to-valley ratio of 4-mm rods aligned perpendicular to the detector panels. The peak and valley positions are indicated by red and yellow dots, respectively, in the bottom left illustration. The image noise was evaluated by the percentage standard deviation of the region of interest (ROI) shown in the uniform area.

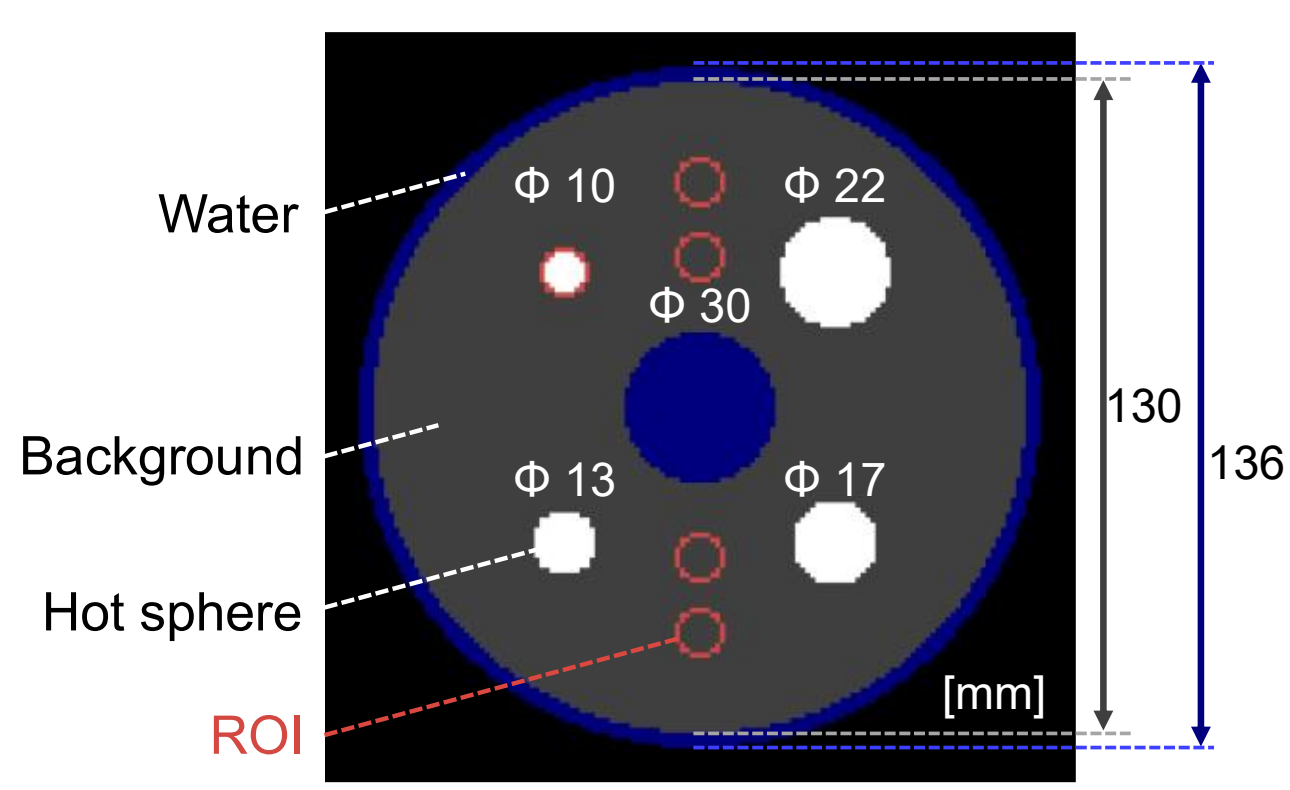

**Fig. 3.** Schematic of an image quality phantom. Circular ROIs with a 10 mm diameter were set in the 10 mm hot sphere and the background area.

### *E. Performance Evaluation*

The image quality of the reconstructed images was evaluated for spatial resolution, image noise, and contrast recovery. Spatial resolution in the multi-rod and uniform phantom was evaluated from the resolution of the 4 mm diameter rods aligned perpendicular to the detector panels (Fig. 2). Percent peak-to-valley ratio (%P2V), which represents the rod resolution, was defined as

$$\%P2V = \left(1 - avg_{v2p}\right) \cdot 100, \tag{3}$$

where $avg_{v2p}$ is the average value of the four valley-to-peak ratios calculated for pairs of rod center points (peak) and midpoints (valley) adjacent to each other. The peak and valley positions are indicated by red and yellow dots, respectively, in the bottom left illustration in Fig. 2. Image noise for the multi-rod and uniform phantom was evaluated from the variation of pixel values in a circular region of interest (ROI). The circular ROI with a 30 mm diameter was set in a slice half the height of the cylindrical uniform area (Fig. 2). Percent standard deviation (%SD) was defined as

$$\%SD = \frac{std_{ROI}}{avg_{ROI}} \cdot 100, \tag{4}$$

where $avg_{ROI}$ and $std_{ROI}$ are the average and standard deviation of the pixel values in the circular ROI. To evaluate the background noise for the image quality phantom, four circular ROIs with a 10 mm diameter were set in the background area of a central slice, +10 mm, +5 mm, -5 mm, and -10 mm slices on either side of the central slice (Fig. 3). Percent background variability (%BV) was defined as

$$\%BV = \frac{\sqrt{\frac{\sum_{i=1}^{n}\left(avg_{BGROI_i} - \frac{1}{n}\sum_{i=1}^{n} avg_{BGROI_i}\right)^2}{n-1}}}{\frac{1}{n}\sum_{i=1}^{n} avg_{BGROI_i}} \cdot 100, \tag{5}$$

where $avg_{BGROI_i}$ is the average of the pixel values in the $i$-th circular background ROI, and $n$ is the number of background ROIs, which is 20. Contrast recovery was evaluated from percent contrast (%contrast), which was defined as

$$\%contrast = \frac{\frac{avg_{HROI}}{\frac{1}{n}\sum_{i=1}^{n} avg_{BGROI_i}} - 1}{\frac{a_H}{a_{BG}} - 1} \cdot 100, \tag{6}$$

where $avg_{HROI}$ is the average of the pixel values in an ROI with a 10 mm diameter in the 10 mm hot sphere, and $a_H$ and $a_{BG}$ were the activity concentration of the hot spheres and background area, respectively.

We investigated the imaging characteristics by varying the TOF resolution, RDE, and image reconstruction method.

#### *1) Spatial resolution and image noise for various TOF resolutions*

We evaluated the spatial resolution and image noise of the dual-panel PET with various TOF resolutions from 30 ps FWHM, which is the current best TOF resolution [5], to 90 ps FWHM. The multi-rod and uniform phantom was used. The data of the non-TOF ring PET were used as a reference for the spatial resolution. The image quality of the MLEM images changes depending on the number of iterations. Thus, we chose the number of iterations with the %SD closest to 10% for the spatial resolution study. For the image noise study, the %P2V was fixed at a value when %SD was 10% in the non-TOF ring PET. The maximum number of MLEM iterations was 200.

#### *2) Image noise for various RDEs*

Reduced Cherenkov photon detection efficiency may cancel the TOF gain in image quality. Thus, we evaluated the image noise in the case of reduced detection efficiency using the multi-rod and uniform phantom. We varied RDE from 10% to 100% at 10% intervals because the valid event ratio where Cherenkov photons trigger both photosensors depends on the measurement setup [24], [28]. In this comparison, we varied the TOF resolution as 30, 40, and 50 ps.

#### *3) Backprojection vs. MLEM methods*

To investigate the advantage of iterative reconstruction in ultrafast TOF resolution, we compared MLEM images with backprojection images using the data of the multi-rod and uniform phantom. For this comparison, the TOF resolution was set to 30 ps. We used the backprojector of the TOF response function and most likely point for the backprojection.

#### *4) Image quality for various TOF resolutions*

The image quality phantom was used to evaluate contrast recovery and image noise. The MLEM was used to reconstruct the data of image quality phantom with 40% RDE [24], [28]. Attenuation correction was applied. The Gaussian post-filtering with 2.0 mm FWHM was applied to suppress the image noise. The maximum iteration was 100, and the number of iterations with the %BV closest to 6% was chosen for visual evaluation.

## III. RESULTS

### *A. Spatial resolution and image noise for various TOF resolutions*

Fig. 4 shows the %P2V-%SD curves in the MLEM iterations with 30, 40, 50, 60, and 90 ps TOF resolutions. The %P2V was increased with improving TOF resolution.

Fig. 5 shows the reconstructed images of the multi-rod area with 30, 40, 50, 60, and 90 ps TOF resolutions in the dual-panel PET and without TOF resolution in the ring PET. As the TOF resolution was improved, the artifacts in the vertical direction were reduced. The 3 mm rods were resolved in the dual-panel PET images with better TOF resolutions than 50 ps, which was the same rod resolution as the non-TOF ring PET. The 4 mm rods were resolved even at 60 ps TOF resolution.

Fig. 6 shows the %P2V of the reconstructed images in Fig. 5. In the dual-panel PET, the %P2Vs at TOF resolutions of 30 ps (99.4%) and 40 ps (97.6%) were higher than in the non-TOF ring PET (92.3%). The average and standard deviation of %SD of the MLEM images were 10.0±0.02%.

Fig. 7 shows the uniform area of the reconstructed images. To compare the noise level at the same spatial resolution, the %P2V of these images was fixed at 92.3%, which was the value when the %SD was 10% in the non-TOF ring PET. As a result, the noise was reduced as the TOF resolution was improved. The %SD at 30, 40, and 50 ps TOF resolutions were 3.5%, 6.8%, and 12.7%, respectively.

### *B. Image noise for various RDEs*

Fig. 8 shows the %SD for various RDEs. The MLEM iterations were selected when %P2V was higher than 91.3% and closest to 92.3%, which was when %SD was 10% in the non-TOF ring PET. The average and standard deviation of %P2V of the MLEM images were 92.3±0.2%. Even when reducing RDE to 20%, the %SD was comparable at 30 ps TOF resolution (6.9%) to that at 40 ps TOF resolution with complete efficiency (6.8%) (gray dashed line). However, the %SD at 30 ps TOF resolution with 10% RDE (8.3%) was worse than that at 40 ps TOF resolution with 50% RDE (7.5%). The %SD at 40 ps TOF resolution with 20% RDE (12.4%) was comparable to that at 50 ps TOF resolution with complete efficiency (12.7%).

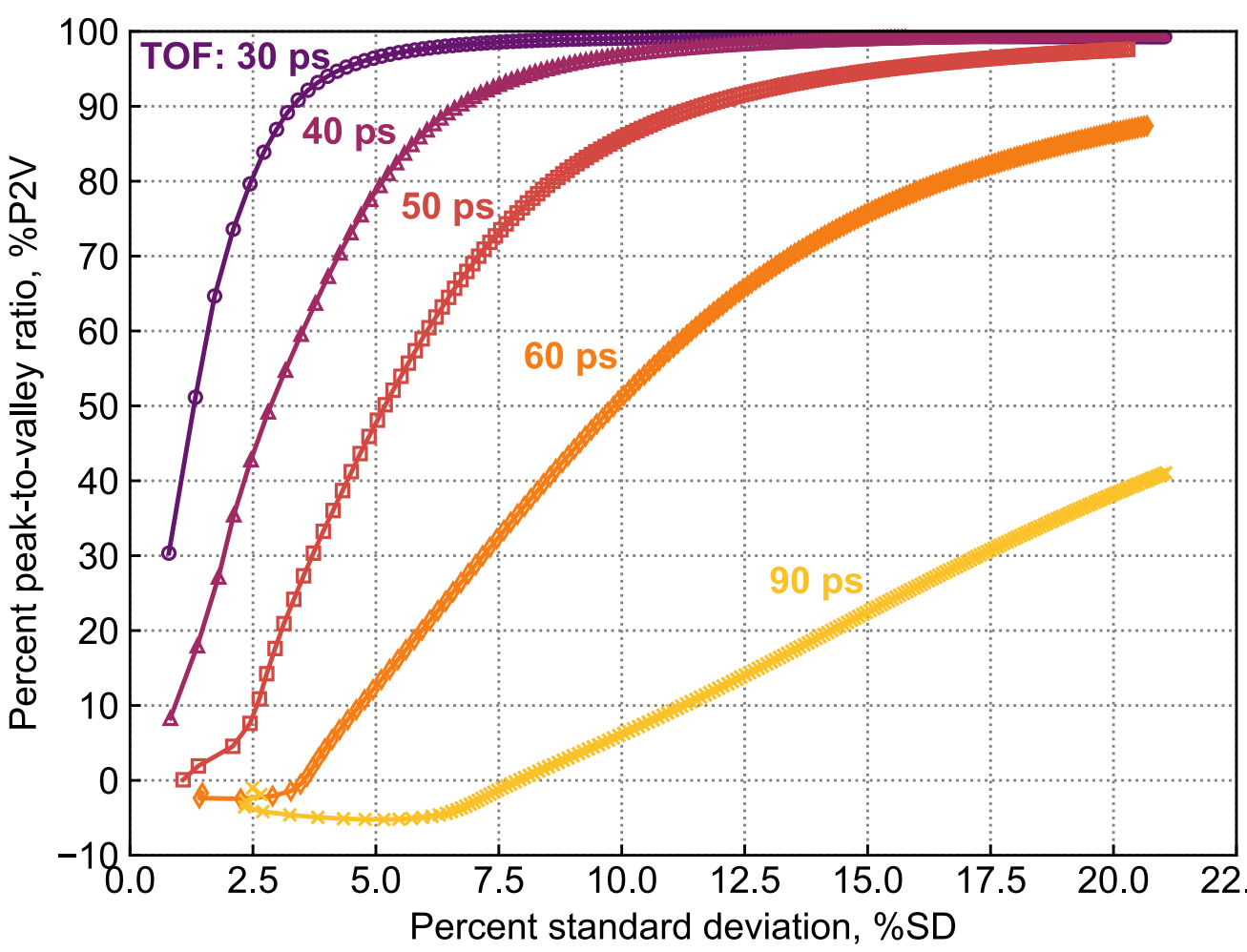

**Fig. 4.** The %P2V-%SD curves in the MLEM iterations with five TOF resolutions. Markers are plotted from one to 200 (left to right). %SD were decreased and %P2V were increased according to the TOF resolution.

### *C. Backprojection vs. MLEM methods*

Fig. 9 shows the %P2V-%SD plot of the MLEM iterations and the backprojection images with the dual-panel PET at 30 ps TOF resolution. The backprojection image had a worse %SD but better %P2V with the most likely point than the TOF response function had. At the fixed %SD, the MLEM images exhibited higher %P2Vs than the backprojection images.

Fig. 10 shows the multi-rod area (top) and uniform area (bottom) of the reconstructed images plotted in Fig. 9. In terms of spatial resolution, the backprojection image obtained with the most likely point approach was more blurred than the MLEM image at the same %SD. In the backprojection images, the rods with diameter larger than 3 mm were resolved with the most likely point, whereas the rods with diameter larger than 4 mm were resolved with the TOF response function.

### *D. Image quality for various TOF resolutions*

Fig. 11 shows the %contrast and %BV curves in the MLEM iterations with 30, 40, 50, 60, and 90 ps TOF resolutions. The %contrast was improved according to the TOF resolution at a fixed %BV.

Fig. 12 shows the reconstructed images of the image quality phantom with 30, 40, 50, 60, and 90 ps TOF resolutions in the dual-panel PET. Three or four iterations were required to have the %BV closest to 6%. As the TOF resolution improved, the artifacts of the hot spheres in the vertical direction were reduced, and the %contrast was improved.

## IV. Discussion

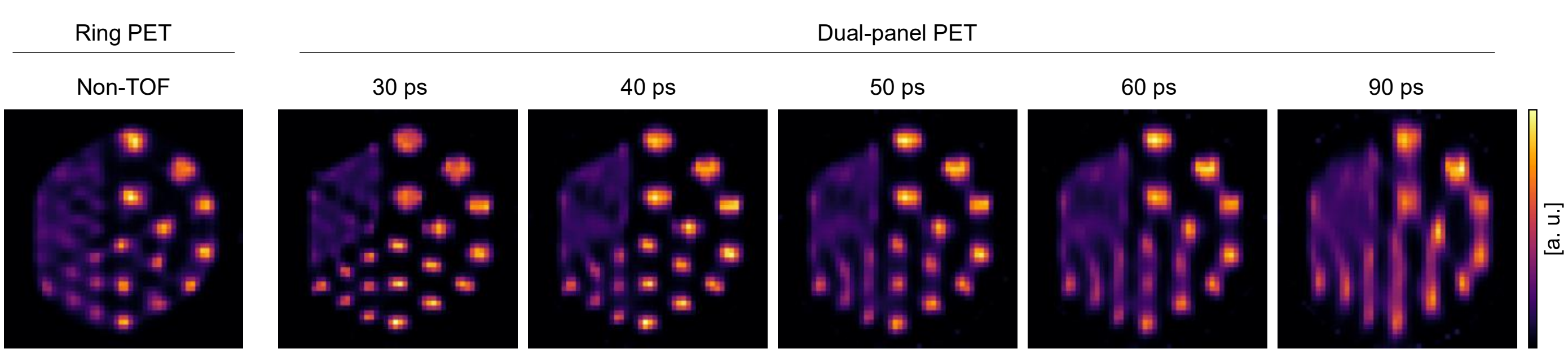

**Fig. 5.** MLEM images of the multi-rod area with various TOF resolutions from 30 ps to 90 ps. The MLEM image of the non-TOF ring PET is shown on the left. The iterations were selected at 10% of the %SD. As the TOF resolution was improved, the artifacts in the vertical direction were reduced in the dual-panel PET. The 3 mm rods were resolved with a better TOF resolution than 50 ps, which was the same rod resolution as that of the non-TOF ring PET.

In terms of the spatial resolution, 3 mm rods were resolved even at 50 ps TOF resolution in the proposed dual-panel PET (Fig. 5), and 40 ps TOF resolution was required to have a similar spatial resolution in the direction perpendicular to the detector panels as for the non-TOF ring PET (Fig. 6). Moreover, the spatial resolution with 30 ps TOF resolution was better than that of the non-TOF ring PET. A simulation study showed the potential of ultrafast TOF to mitigate the spatial resolution limit using a ring geometry [40]. The present result showed the same advantage of ultrafast TOF when using the dual-panel geometry. The diameter of the ring PET and the panel-to-panel distance of the dual-panel PET were 300 mm in this comparison. The 300 mm diameter of the ring PET is insufficient for human whole-body imaging, and conventional whole-body PETs have larger diameters than 700 mm [4], [6]. The large ring diameter degrades the spatial resolution because of the photon non-colinearity. On the other hand, the dual-panel configuration can bring the detector panels close together. When the bottom detector panel was embedded in the patient's bed, we believe human whole-body imaging is possible with a 300 mm panel-to-panel distance. The required TOF resolution for the dual-panel PET in the present setup can be mitigated from 40 ps to have the same spatial resolution as the 800 mm-diameter ring PET. The rods 3 mm in diameter were resolved even with 40 ps TOF resolution, corresponding to 6.0 mm spatial resolution. This rod resolution was possible because the oblique LORs compensated for the spatial information in the direction perpendicular to the detector panels.

The image noise was reduced according to the TOF resolution when the same efficiency was maintained in the detectors (Fig. 7). In the case of reduced efficiency of the detectors (Fig. 8), image noise was comparable at 30 ps TOF resolution with 20% RDE and at 40 ps TOF resolution with complete efficiency. We showed that the TOF information could compensate for the loss of image quality due to reduced efficiency. The detection efficiency of the Cherenkov event is determined by multiple detector parameters, such as the stopping power of the BGO crystals, the number of the generated photons that reach the photocathode, and the event rate triggered by the Cherenkov photons [24]. We simplified these factors into one parameter, RDE. Further study is needed to investigate how these specific detector parameters change the detection efficiency.

The MLEM image showed better spatial resolution than the backprojection images did for the same noise level (Fig.9, Fig. 10). This result showed that the iterative reconstructions improved the image quality even with the ultrafast TOF resolution. Although MLEM requires iterations, only three iterations were required to obtain the same %P2V as backprojection with a better %SD, and only four iterations were required to obtain a better %P2V with the same %SD, which were not time-consuming.

The %contrast of the 10 mm hot sphere was improved according to the TOF resolution with the fixed %BV of 6% (Fig. 12). The benefit of TOF in terms of contrast and image noise was shown with a plausible RDE of 40 %[24], [28]. Fast TOF resolutions also contributed to the fast convergence of the MLEM, which enabled the reconstructed images to be obtained with a good balance of contrast and noise in three or four iterations. The ordered subset expectation maximization (OSEM) will further reduce the reconstruction time [41].

The assumptions in timing performance and geometry of the detector were based on the CRI-MCP-PMT with 30 ps TOF resolution [5] and the commercially available multi-anode MCP-PMT (R10754-07-M16, Hamamatsu Photonics K. K.), respectively. The results in the present study showed the necessity of developing the BGO-integrated multi-anode MCP-PMT. The detector design to realize the proposed dual-panel PET will be considered using optical simulation in future work.

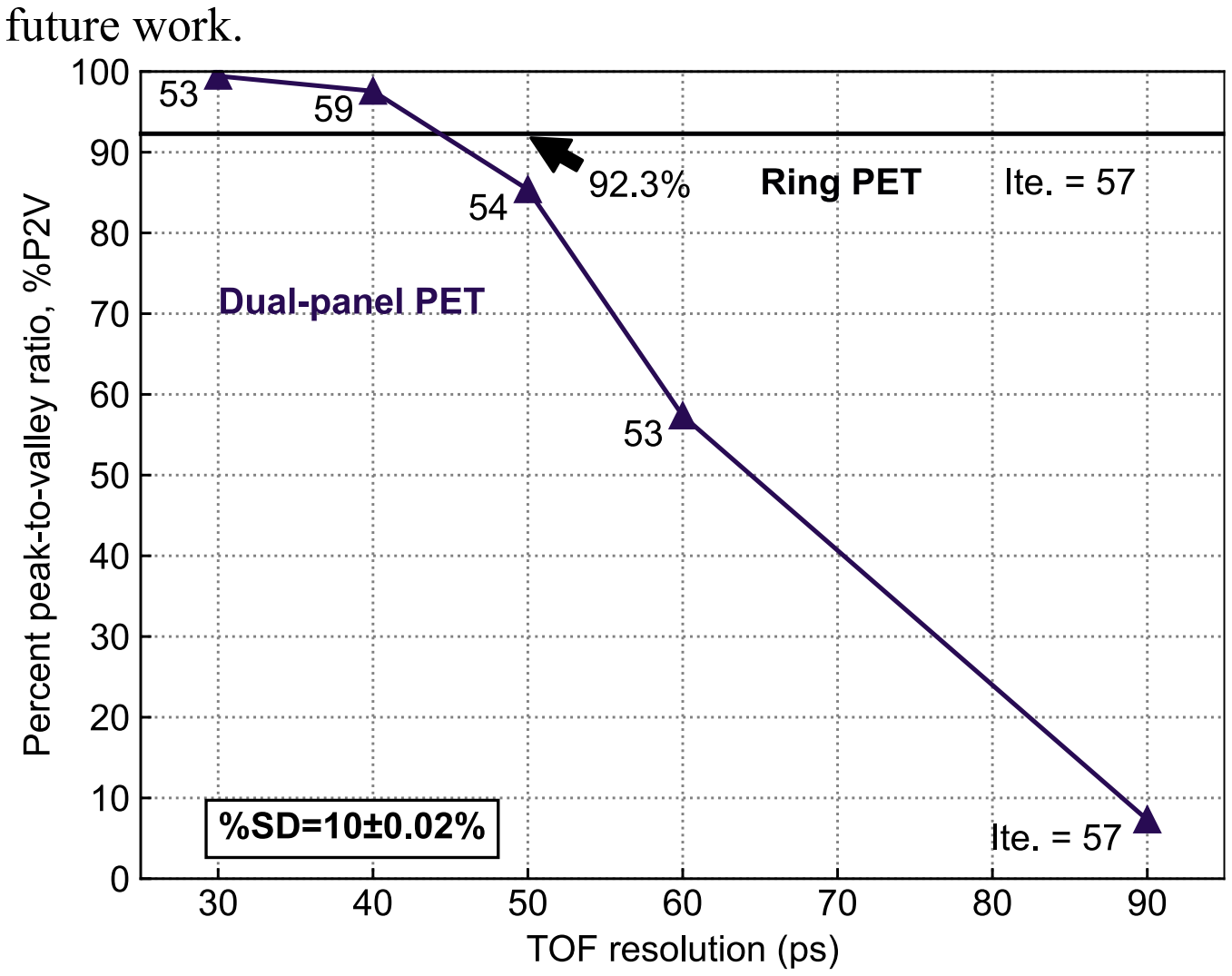

**Fig. 6.** Relation between %P2V and TOF resolution. The numbers of MLEM iterations are shown below each symbol for the dual-panel PET and below the line for the ring PET. The %P2V was higher at 30 ps TOF resolution (99.4%) and 40 ps TOF resolution (97.6%) in the dual-panel PET than the non-TOF ring PET (92.3%).

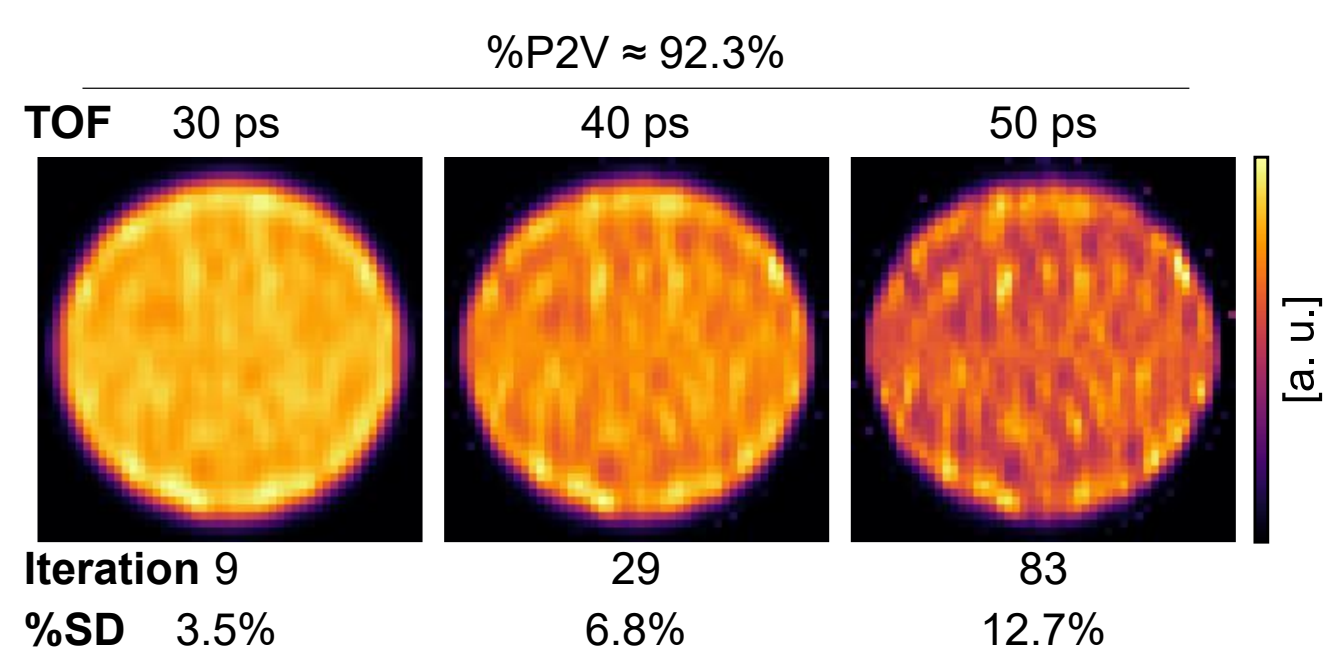

**Fig. 7.** MLEM images of the uniform area for TOF resolutions from 30 ps to 50 ps. The iterations were selected for a %P2V of 92.3% when %SD was 10% in the non-TOF ring PET. The image noise was decreased as the TOF resolution was improved.

In this study, we modeled the pixelized BGO crystals as the MCP-PMT window. However, it likely needs to be a monolithic crystal to maintain the vacuum inside the MCP-PMTs in realistic situations. Monolithic crystals will show better positional resolution than pixelized crystals having the same pitch as the anode pitch, because of the anger calculation [42]. On the other hand, the uniformity of the positional resolution will be degraded. TOF resolution will also vary [43]. These factors should be considered in future work.

The TOF-PET detector with BGO crystals shows a long tail in the timing histogram [44]. This tail is due to the variance of the number of the detected Cherenkov photons or the contamination of events triggered by only the scintillation photons. Several research groups are developing methods to identify whether single events are triggered by the Cherenkov photons or scintillation photons [24], [25], [27], [28], [44]. In this study, we assumed that this identification could be done correctly. The impact of identification failures on image quality needs to be investigated in the future.

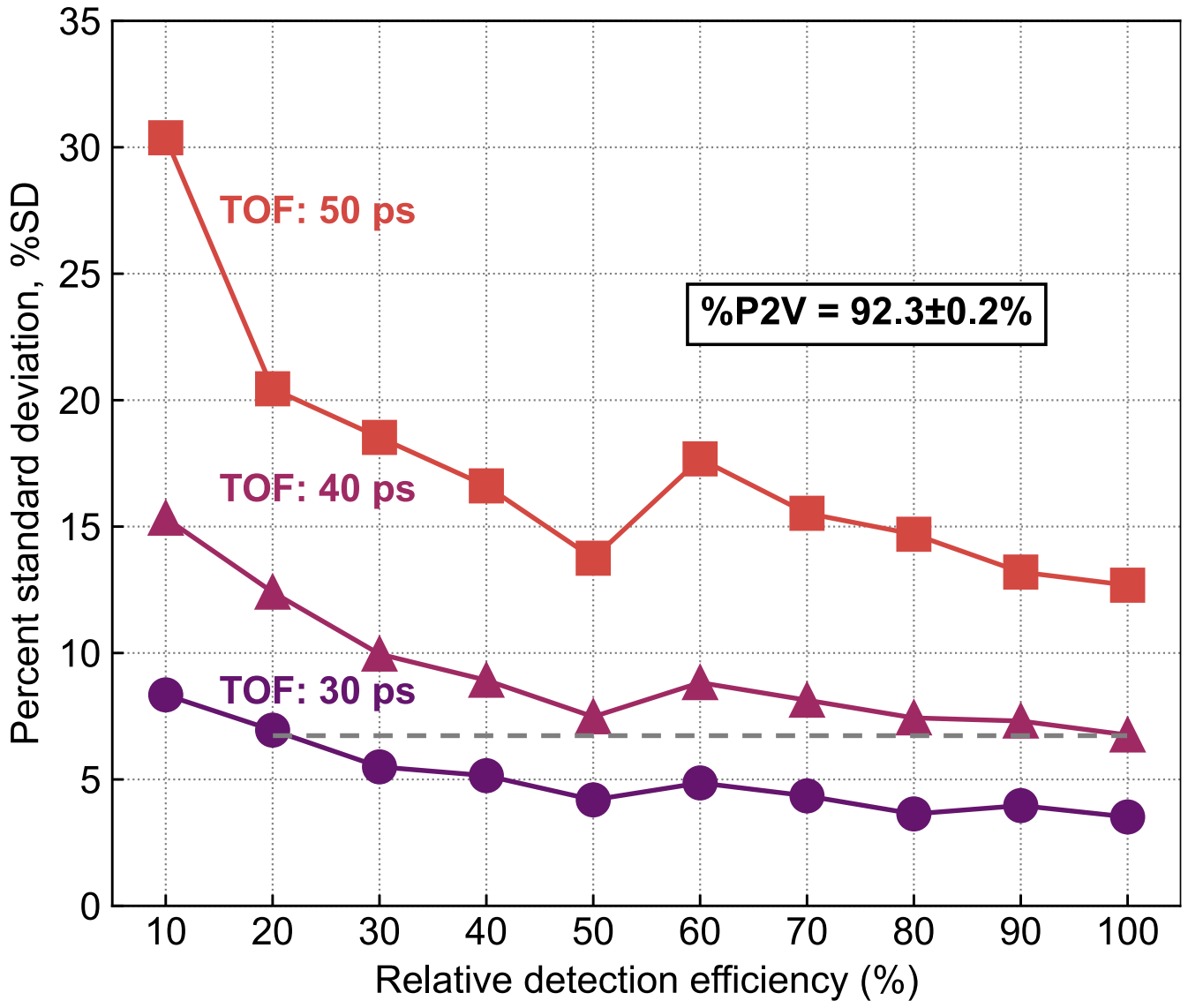

**Fig. 8.** Relation between RDE and %SD. The MLEM iterations were selected when %P2V was higher than 91.3% and closest to 92.3%, which was when %SD was 10% in the non-TOF ring PET. Even with RDE reduced to 20%, the %SD at TOF resolution of 30 ps (6.9%) was better than that at 40 ps with complete efficiency (6.8%) (gray dashed line). The %SD at 30 ps with 10% RDE (8.3%) was worse than that at 40 ps with 50% RDE (7.5%). The %SD at 40 ps with the 20% RDE (12.4%) was comparable to that at 50 ps with complete efficiency (12.7%).

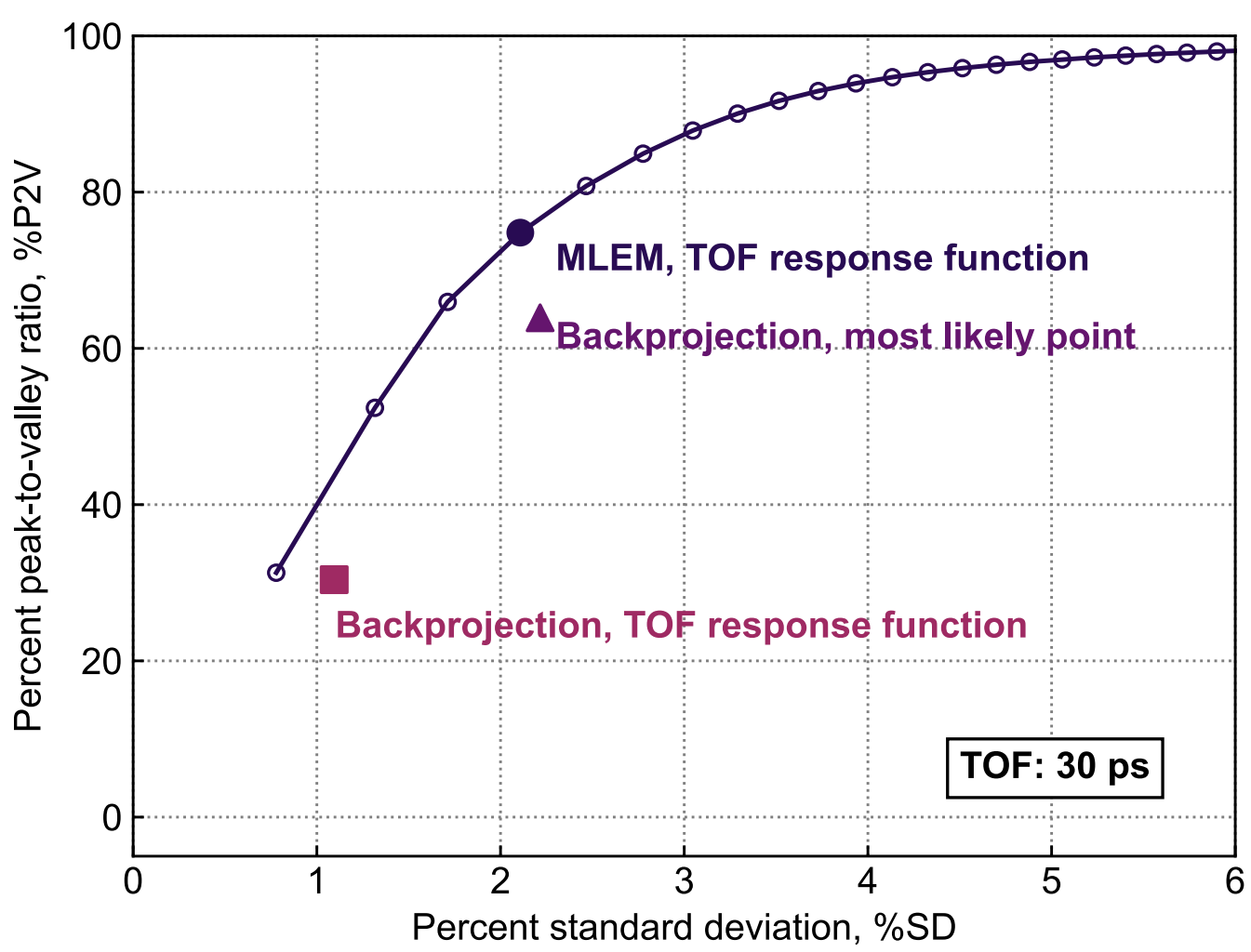

**Fig. 9.** The %P2V-%SD plot of MLEM iterations and backprojection images with the dual-panel PET at 30 ps TOF resolution. Markers are plotted from one to 22 (left to right) in MLEM plot. In the backprojection images, the most likely point had a worse %SD but better %P2V than the TOF response function. At a fixed %SD, the MLEM images showed higher %P2Vs than the backprojection images.

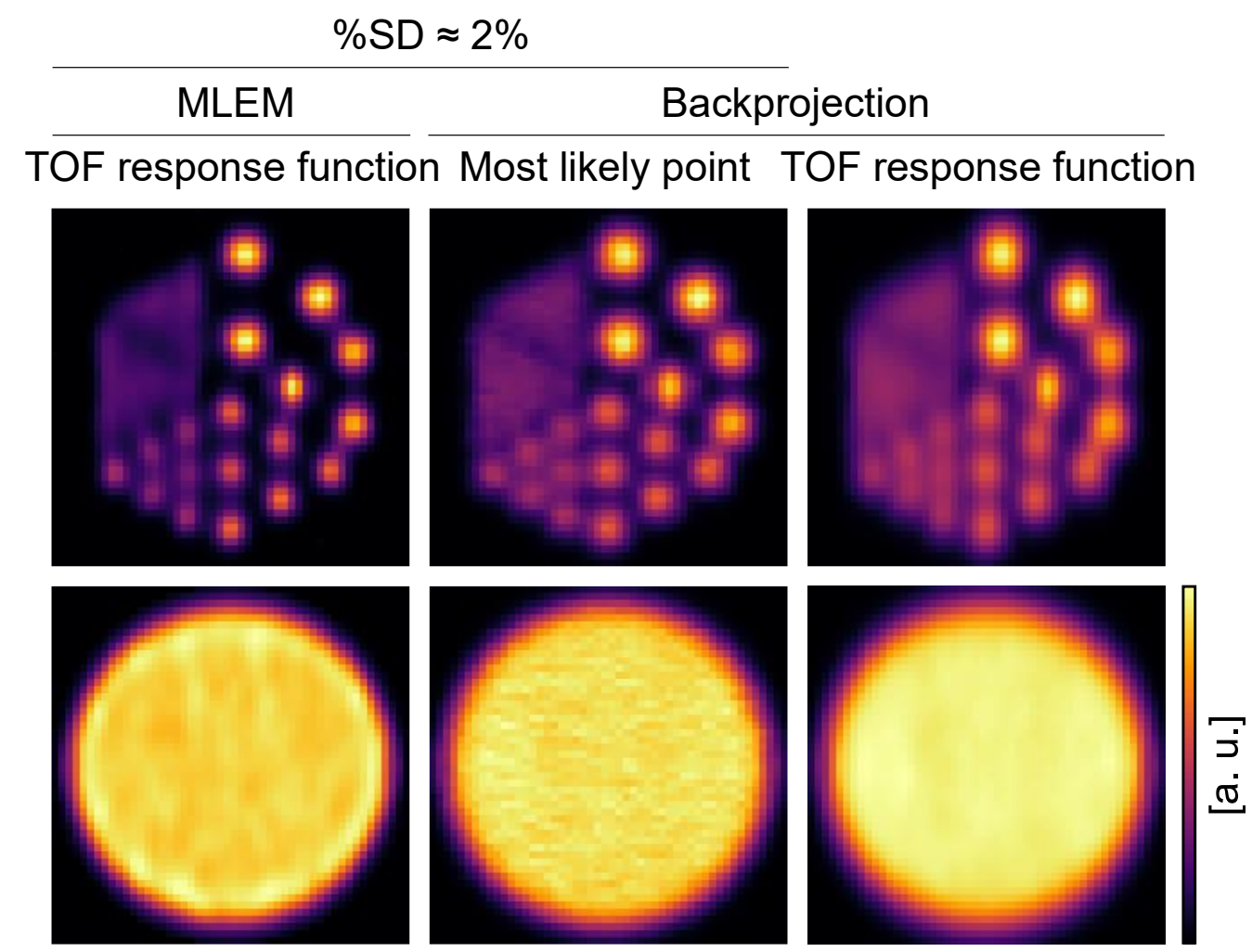

**Fig. 10.** Comparison of the reconstructed images between MLEM and backprojection with the dual-panel PET at 30 ps TOF resolution. The multi-rod areas and uniform areas were shown in the top row and bottom row, respectively. The MLEM image showed better spatial resolution than the backprojection image with the TOF response function at the same image noise level. In the backprojection images, the rods larger than 3 mm diameter were resolved with the most likely point, whereas the rods larger than 4 mm were resolved with the TOF response function.

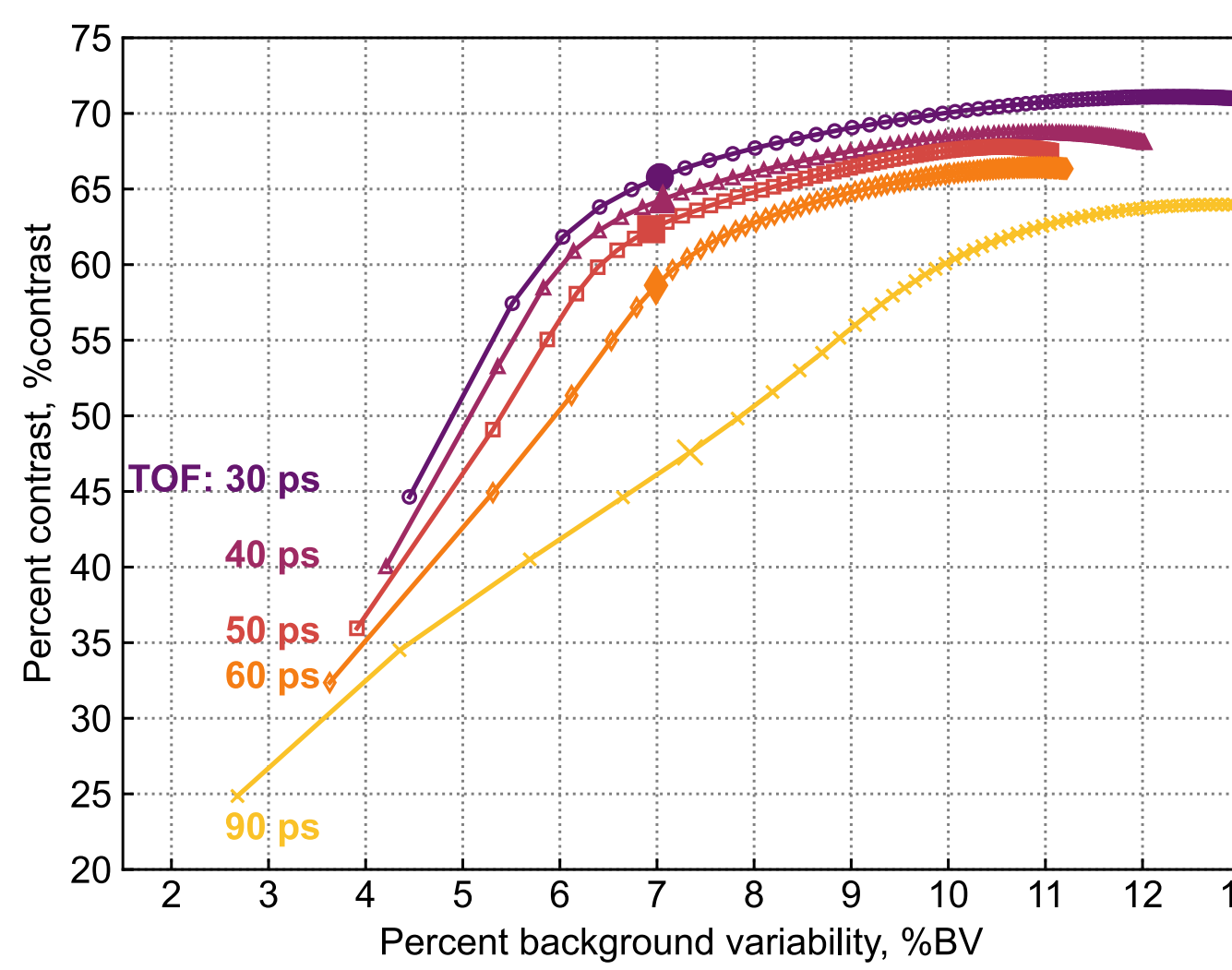

**Fig. 11.** The %contrast and %BV curves in the MLEM iterations with five TOF resolutions. Markers are plotted for each iteration (left to right). The %contrast was improved according to the TOF resolution at a fixed %BV.

Razdevšek *et al.* [14] investigated the imaging performance of a similar dual-panel system using Monte Carlo simulation. They used lutetium oxyorthosilicate crystals as scintillators. On the other hand, we used BGO crystals for the hybrid detection of Cherenkov and scintillation photons, which was the major difference between this study and the previous study.

As a reference scanner for the spatial resolution, we used BGO crystals as scintillators of 5 mm thickness for the non-TOF ring PET. However, Lu-based scintillators with 20 mm thickness are used for a commercial scanner having 214 ps TOF resolution [4]. In addition to the TOF information, the higher stopping power contributes to better image noise levels than the results obtained in this study. We need to investigate the required specifications for the dual-panel PET to outperform the Lu-based ring PET in a further study.

## V. Conclusion

We investigated the imaging characteristics of the dual-panel PET with compact detector panels and ultrafast TOF resolution for the PET-guided surgery and beam monitoring applications using Geant4 simulation. The detector panel (137 × 137 mm$^2$) consisted of 5.0 mm-thick BGO pixelized crystals with a 5.75 mm pitch, and the two panels were separated by 300 mm. The TOF resolution required for the dual-panel PET to have a similar spatial resolution performance to that of the non-TOF ring PET was determined to be 40 ps for the same efficiency in the detectors. In the case of half the efficiency, we showed that the TOF resolution should be 30 ps to maintain a similar level of image noise to that of the 40 ps TOF resolution with complete efficiency. The MLEM images showed better spatial resolution than the backprojection images at the same noise level, even at 30 ps TOF resolution. TOF gain improved the contrast and image noise at 40% RDE in image quality evaluation. In conclusion, our simulation results supported the feasibility of the proposed dual-panel PET using compact detector panels with the ultrafast TOF detector.

## Acknowledgment

All authors declare that they have no known conflicts of interest in terms of competing financial interests or personal relationships that could have an influence or are relevant to the work reported in this paper.

This work was partially supported by the Japan Society for the Promotion of Science (JSPS) KAKENHI Grant Number 20H05667 and the Nakatani Foundation.

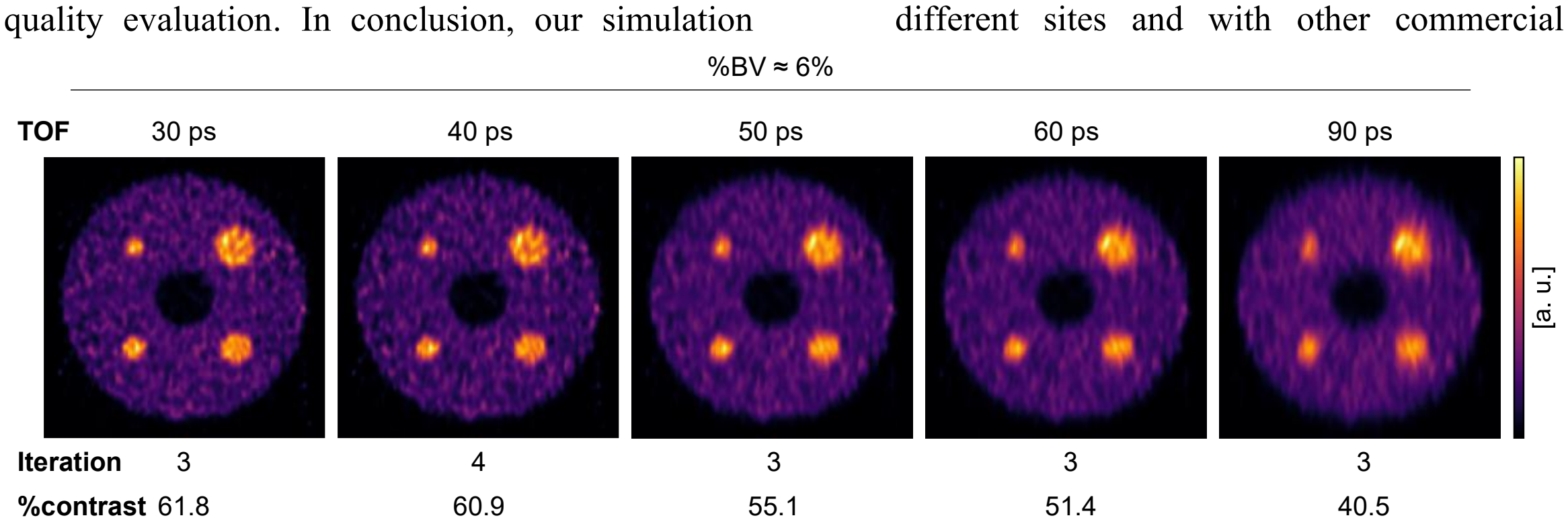

**Fig. 12.** MLEM images of the image quality phantom with various TOF resolutions from 30 ps to 90 ps in 40% RDE. Three or four iterations were required to have the %BV closest to 6%. As the TOF resolution improved, the artifacts of the hot spheres in the vertical direction were reduced, and the %contrast was improved.